\documentclass[twocolumn,prl,amsmath]{revtex4}
\usepackage[dvips]{graphicx}
\usepackage{pstricks,psfig,epsfig}
\usepackage{amsmath}
\usepackage{xspace}

\newcommand{\eg}{{\it e.g.}\xspace}

\newcommand{\ave}[1]{\left\langle#1 \right\rangle}

\newcommand{\prepsection}[1]{{\it #1} --- }
\renewcommand{\section}[1]{\prepsection{#1}}

\newcommand{\elabel}[1]{\label{eq:#1}}
\newcommand{\eref}[1]{Eq.~(\ref{eq:#1})}

\newcommand{\flabel}[1]{\label{fig:#1}}

\begin{document}
\title{Critical phenomena in atmospheric precipitation}
\author{Ole Peters}
\email{ole.peters@physics.org}
\homepage{http://www.santafe.edu/~ole}
\affiliation{CNLS, Los Alamos National Laboratory, MS-B258, Los Alamos, NM 87545, USA.\\
Santa Fe Institute, 1399 Hyde Park Road, Santa Fe, NM 87501, USA.\\
Department of Atmospheric Sciences  and
 Institute of Geophysics and Planetary Physics, 
 University of California, Los Angeles, 405 Hilgard Ave., Los Angeles, CA 90095-1565, USA.
}

\author{J.~David Neelin}
\affiliation{Department of Atmospheric Sciences  and
 Institute of Geophysics and Planetary Physics, 
 University of California, Los Angeles, 405 Hilgard Ave., Los Angeles, CA 90095-1565, USA.}

\begin{abstract}
Critical phenomena near continuous phase transitions are typically
observed on the scale of wavelengths of visible
light\cite{KleinTisza1949}. Here we report similar phenomena for
atmospheric precipitation on scales of tens of kilometers. Our
observations have important implications not only for meteorology but
also for the interpretation of self-organized criticality (SOC) in
terms of absorbing-state phase transitions, where feedback mechanisms
between order- and tuning-parameter lead to
criticality.\cite{DickmanVespignaniZapperi1998} While numerically the
corresponding phase transitions have been
studied,\cite{DickmanETAL2001,ChristensenETAL2004} we characterise for
the first time a physical system believed to display
SOC\cite{PetersHertleinChristensen2002} in terms of its underlying
phase transition. In meteorology the term quasi-equilibrium
(QE)\cite{ArakawaSchubert74} refers to a state towards which the
atmosphere is driven by slow large-scale processes and rapid
convective buoyancy release. We present evidence here that QE,
postulated two decades earlier than SOC\cite{BakTangWiesenfeld1987},
is associated with the critical point of a continuous phase transition
and is thus an instance of SOC.\vspace{0.5cm}

Journal reference: Nature Physics {\bf 2}, 393 - 396 (2006). doi:10.1038/nphys314
\end{abstract}
\maketitle

Self-organized criticality has been proposed as an explanation for
scale-free behaviour in many different physical
systems\cite{ChristensenMoloney2005}. In most of these, however, it is
impossible to measure standard observables for critical phenomena,
such as order parameters, tuning parameters, or
susceptibilities. Consequently, despite theoretical
advances\cite{DickmanVespignaniZapperi1998,DickmanETAL2001,ChristensenETAL2004},
SOC has only loosely been connected to the broader field of critical
phenomena. The present study helps position SOC as a sub-branch of
critical phenomena by examining a system where the identification and
measurement of standard observables is feasible.

At short time scales the majority of tropical rainfall occurs in
intense rain events that exceed the climatological mean rate by an
order of magnitude or more. Precipitation has been found to be
sensitive to variations in water vapour along the vertical on large
space and time scales both in
observations\cite{BrethertonPetersBack04,Parsonsetal2000} and in
models.\cite{Tompkins01,Grabowski03,Derbyshireetal05} This is due to
the effect of water vapour on the buoyancy of cloud plumes as they
entrain surrounding air by turbulent mixing. We conjecture that the
transition to intense convection, accompanying the onset of intense
precipitation, shows signs of a continuous phase transition. The water
vapour, $w$, plays the role of a tuning parameter and the
precipitation rate, $P(w)$, is the order parameter
Note that such a large-scale continuous phase transition involving the
flow regime of the convecting fluid is entirely different from the
well-known discontinuous phase transition of condensation at the
droplet scale.  
We analyzed satellite microwave retrievals of rainfall, $P$, water
vapour, $w$, cloud liquid water and sea surface temperature (SST) from
the Tropical Rainfall Measuring Mission from 2000 to 2005.
Observations from the western Pacific provided initial support for our
conjecture: 
a power-law pick up of the order parameter above a critical value of
the tuning parameter, $w_c$, was observed. We proceded to test whether
other observables also behaved as predicted by the theory of phase
transitions.

As motivation for our conjecture consider a generic lattice-based
model which exhibits a continuous phase
transition. Particle-conserving rules defining the model ascribe a
number of particles to every lattice site, and demand hopping of
particles to nearest-neighbour sites when a local density threshold is
exceeded. The global effect of these rules is a phase transition at a
critical value of the global particle density between a quiescent
phase (where the system eventually settles into a stable
configuration) and an active phase (where stable configurations are
inaccessible). The tuning parameter is the particle density and the
order parameter is identified as the density of active
sites\cite{MarroDickman1999}.

SOC can be described in terms of such absorbing-state phase
transitions.\cite{DickmanVespignaniZapperi1998,MarroDickman1999}
Here a coupling between order parameter and tuning parameter is
introduced by opening the boundaries and adding a slow drive: whenever
activity ceases, a new particle is added to the system, i.e., an
increase in the tuning parameter. Large activity on the other hand
leads to dissipation (particle loss) at the boundaries, i.e., a
reduction of the tuning parameter. Such open, slowly driven systems
organise themselves to the critical point of the corresponding (closed
boundaries, no drive) absorbing state phase transition. The critical
behaviour as derived from finite-size scaling analyses is the same in
both cases\cite{DickmanETAL2001,ChristensenETAL2004}, although the
reason for this universality is not fully
understood.\cite{PruessnerPeters2005} The scale-free avalanche size
distributions in SOC models result from the proximity of the system to
a critical point.

From the meteorological perspective, a related motivation for our
conjecture arises. Atmospheric convection has long been viewed
similarly in terms of a slow drive (surface heating and evaporation)
and fast dissipation (of buoyancy and rainwater) in precipitating
convection. Surface heating and evaporation drive turbulent mixing
that maintains a moist atmospheric boundary layer. Combined with
radiative cooling, conditional instability is created---while
sub-saturated air remains stable, saturated condensing plumes can rise
through the full depth of the tropical troposphere. The fast
dissipation by moist convection prevents the troposphere from
deviating strongly from marginal stability.\cite{XuEmanuel89} Although
observational tests of this approximate QE state of the tropical
troposphere have limited precision, it forms the basis of most
convective parameterizations in large scale models\cite{Arakawa04} and
much tropical dynamical theory.\cite{EmanuelNB94,NeelinZeng00}
Taking large-scale flows into account modifies the process in space
and time but does not change it fundamentally. This perspective
suggests that a critical point in the water vapour would act as an
attractor. Indeed this is basically the convective QE
postulate.\cite{ArakawaSchubert74}

The critical value $w_c$ depends, \eg, on atmospheric temperature, but
for present purposes this translates well enough into a critical
amount of water vapour for a given climatic region. Regions here are
defined by longitude ranges given in the caption of Fig.~1
corresponding to major ocean basins, for oceanic grid-points within
20S-20N. Data are collected at 0.25 degree latitude-longitude
resolution. The observable $w$ captures vertically integrated, or
column, water vapour. It is given as a volume per area in units of mm.

In Fig.~1 we show as a function of the tuning parameter $w$ the
average value of the order parameter $\ave{P}(w)$ and the
susceptibility of the system, represented by the order parameter
variance, $\sigma^2_P(w)$, discussed following \eref{susceptibility}.
The ensemble size for the average ranges from a few thousand at
extremes to $10^6$ at typical $w$-values. Above $w_c$, the order
parameter is well approximated by the standard form\cite{Yeomans1992}
\begin{equation}
\ave{P}(w)=a(w-w_c)^{\beta}, 
\elabel{order} 
\end{equation}
where $a$ is a system-dependent constant and $\beta$ is a universal
exponent. The deviations from power-law behaviour below $w_c$ in the
main graph of Fig.~1 are typical of critical systems of finite
size.\cite{FisherBarber1972}

\begin{figure}
 \noindent\includegraphics[width=20pc]{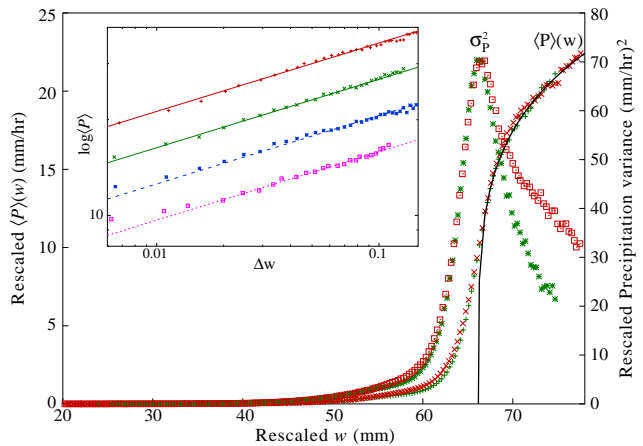}
\caption{{\bf Order parameter and susceptibility.} The main figure
shows the collapsed (see text) precipitation rates $\ave{P}(w)$
 and their variances $\sigma^2_P(w)$
for the tropical Eastern (red) and Western (green) Pacific as well as a
power-law fit above the critical point (solid line). 
The inset displays on double-logarithmic
scales the precipitation rate as a function of reduced water vapour
(see text) for Western Pacific (green, 120E to 170W), Eastern Pacific
(red, 170W to 70W), Atlantic (blue, 70W to 20E), and Indian Ocean
(pink, 30E to 120E). Data are shifted by a small arbitrary factor
for visual ease. The straight lines are to guide the eye. They all
have slope 0.215, fitting the data from all regions well.}
\label{fig:fig1}
\end{figure}

The critical value $w_c$ is non-universal and changes with regional
climatic conditions, as does the amplitude $a$. 
To test the degree to which curves from different regions $i$
collapse, we re-scaled the $w$-values in Fig.~1 by factors $f_w^i$,
reflecting the non-universality of $w_c$ and $\ave{P}(w)$ and
$\sigma^2_P(w)$ by $f_P^i$ and $f_{\sigma^2}^i$, respectively (setting
Western Pacific factors to one). For visual clarity, the data
collapse in Fig.~1 is shown only for the Eastern and Western
Pacific---climatically very different regions. Similar agreement
occurs for other regions (steps in the rescaling and figures for all
regions are provided in the Supplementary Information). The exponent
$\beta$ seems to be universal and independent of the climatic
region. In the inset to Fig.~1 we show the average precipitation as a
function of the reduced water vapour $\Delta w \equiv (w-w_c)/w_c$ in
a double-logarithmic plot.  Importantly, power laws fitted to these
distributions all have the same exponent (slope) to within $\pm 0.02$.
The data points in Fig.~1 represent the entire observational period,
including all observed SSTs. Conditioning averages by SST ranges
yields similar results (see Fig.~3 and Supplementary Information),
reducing the subcritical part of the curves slightly.

We define the susceptibility $\chi(w;L)$ via the variance of the order
parameter $\sigma^2_P$:
\begin{equation}
\chi(w;L) = L^d \sigma^2_P(w;L),
\elabel{susceptibility}
\end{equation}
where $d$ denotes the dimensionality of the system and $L$ the spatial
resolution. Fig.~1 shows a suggestive increase in $\sigma^2_P$
near $w_c$, and indicates that standard methods for critical phenomena
can sensibly be applied. 

Next we test for finite-size scaling. Because our system size cannot
be changed, we identify the spatial data resolution $L$ as the
relevant length scale. Changing $L$ has the effect of taking averages
over different numbers of degrees of freedom and allows one to
investigate the degree of spatial correlation. The finite size scaling
ansatz for the susceptibility is
\begin{equation}
\chi(w;L)=L^{\gamma/\nu}\tilde{\chi}(\Delta w L^{1/\nu}),
\elabel{susceptibility_fss}
\end{equation}
defining $\gamma$ and $\nu$ as the standard critical exponents and the
usual finite-size scaling function $\tilde{\chi}(x)$, constant for
small arguments $|x| \ll 1$ and decaying as $|x|^{-\gamma}$ for large
arguments $|x| \gg 1$.\cite{PrivmanHohenbergAharony1991} The variance
$\sigma^2_P(w;L)$ is affected by uncertainties in $w$ and $w_c$, making
precise quantification of $\chi(w;L)$ difficult. We therefore do not
estimate $\gamma$ from the $w$-dependence of $\chi(w;L)$, corresponding
to large arguments $|x|$ in \eref{susceptibility_fss} but effectively
fix $\Delta w=0=x$ and obtain $\gamma/\nu$ from the $L$-dependence of
the maximum susceptibility $\chi^{\text{max}}(L)$.

The variance of the average $\ave{P}(w;L)$ over $L^d$ independent
degrees of freedom decreases as $\sigma^2_P(w;L) \propto L^{-d}$. In a
critical system, however, the diverging bulk correlation length $\xi
\propto (\Delta w)^{-\nu} \gg L$ (small argument in
\eref{susceptibility_fss}) prohibits the assumption of
independence. In this case \eref{susceptibility_fss} with
\eref{susceptibility} yields
\begin{equation}
\sigma^{2 \text{ max}}_P(L) \propto L^{\gamma/\nu-d}.
\end{equation} 
Coarsening the spatial resolution of the data, we find in Fig.~2
that $\sigma^{2 \text{ max}}_P(L)$ scales roughly as $L^{-\lambda}$, with
$\lambda=0.46(4)$. 
This suggests the exponent ratio $\gamma/\nu=1.54(4)$. 
\begin{figure}
 \noindent\includegraphics[width=20pc]{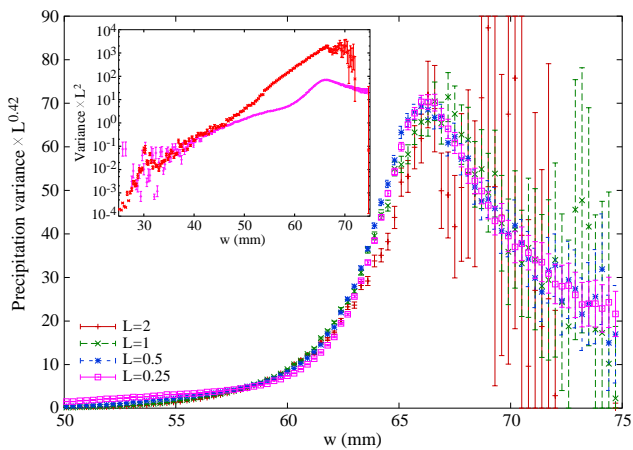}
\caption{{\bf Finite-size scaling.} The variance of the order parameter
$\sigma^2_P(w)$
as a function of $w$, rescaled with $L^{0.42}$ for system sizes
0.25$^{\circ}$, 0.5$^{\circ}$, 1$^{\circ}$, and 2$^{\circ}$ in the
Western Pacific. From $w \approx 57$~mm, this produces a good
collapse. The inset shows that away from the critical point, up to $w
\approx 40$~mm a trivial rescaling with $L^{d=2}$ works
adequately. This suggests that the non-trivial collapse is indeed a
result of criticality.  } \flabel{fig2}
\end{figure}
At criticality, the spatial decay of correlations between order
parameter fluctuations becomes scale-free.\cite{Yeomans1992} This is
equivalent to a non-trivial power-law dependence of the
order-parameter variance on $L$ (see Supplementary Information for
details and conditions). Hence, Fig.~2 indicates a
scale-free correlation function of fluctuations in the rain rate in
the range of 25~km to 200~km. This suggests that the meteorological
features known as mesoscale convective systems\cite{Houze93} are
long-range correlation structures akin to critical
clusters.\cite{StaufferAharony1992} Synoptic inspection indicates that
the high rain rate phase and critical region of Fig.~1 come substantially 
from points within such complexes (examples are provided in the 
Supplementary Information).
\begin{figure}
 \noindent\includegraphics[width=20pc]{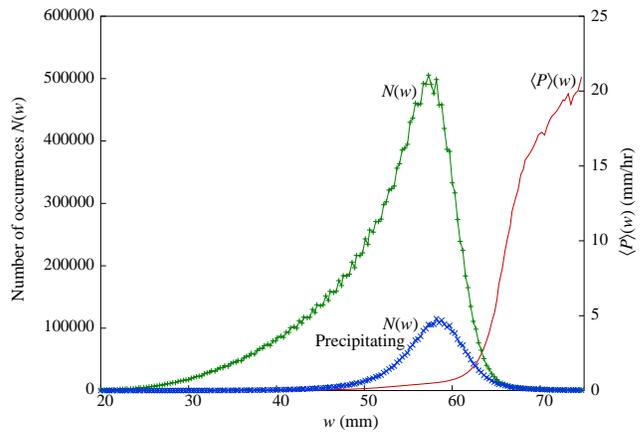}
\caption{{\bf Residence times.} The number of times $N(w)$ an atmospheric
pixel of $0.25^{\circ} \times0.25^{\circ}$ was observed at water
vapour $w$ in the western Pacific, given a sea surface temperature
within a $1^{\circ}$C bin at $30^{\circ}$C. The green and blue lines show 
residence time for all points and precipitating points, respectively.
The red line shows the order-parameter pick-up $\ave{P}(w)$ for orientation 
(precipitation scale on the
right).}  \flabel{fig3}
\end{figure}
The question of self-organisation towards the critical point of the
transition is addressed by displaying the residence times of the
system in Fig.~3. This is the number of observations in the 5-year
period where the system was found at a given level of water vapour. A
slowly driven system would be expected to spend a significant amount
of time in the low-$w$ phase because when it fluctuates into this
phase \eg due to some large-scale event, it takes a long time to
recover. Therefore the distribution decreases slowly towards low
values of $w$. The fast dissipation mechanism, on the other hand,
ensures that the system leaves the high-$w$ regime relatively quickly
when it fluctuates into it. Consequently the distribution decreases
rapidly towards large values of $w$. For the properties of rainfall,
the part of the distribution in Fig.~3 comprised only of observations
with rainfall is of interest, seen as the blue line in Fig.~3. We note
that the system is most likely to be found near the beginning of the
intense precipitation regime. Almost the entire weight of the
distribution of rainy times is concentrated here.

Meteorologically, these results suggest a means to redefine and extend
convective QE, both empirically and theoretically. In its simplest
application QE assumes that the relationship among atmospheric column
thermodynamic variables is pinned close to the point where deep
convection and precipitation set in. Fig.~3 shows this to be a
reasonable first approximation , but it also implies associated
critical phenomena. A loss term $\ave{P}(w)$ of the form of
\eref{order} implies the absence of any well-defined convective time
scale. Scale-free distributions of event
sizes\cite{PetersHertleinChristensen2002} and the spatial correlation
behavior seen in Fig.~2 may result from this proximity to an apparent
continuous phase transition.

These findings beg for a simple SOC-type model of the atmospheric
dynamics
responsible for the critical behaviour. While the physics must conform
with recent cloud-resolving model analysis of mesoscale
aggregation\cite{Tompkins01,BrethertonBK05}, our results point to the
key role of excitatory short-range interactions, essential for
critical phenomena of the type seen here. This study advances our
understanding of SOC as a critical phenomenon, identifying the
underlying phase transition and associated critical phenomena. Beyond
scale-free event size distributions it furnishes direct evidence, for
the first time, for an underlying phase transition in a physical
system.

\section{Methods}
Data are from the TMI (TRMM microwave imager), processed by Remote
Sensing Systems (RSS). The spatial resolution reflects the footprint
of the instrument. As with any satellite retrieval product, it is
necessary to consider whether the algorithm assumptions could impact
the results. The microwave retrieval algorithm is that used on Special
Sensor Microwave Imager (SSM/I) data.\cite{WentzSpencer1998} The
combination of four microwave channels permits independent retrieval of
water vapour and condensed phase water (with SST and surface wind
speed), while an empirical relation is used to partition cloud water
and rain.
Column water vapour validates well against in situ sounding
data, which also show that daily variations are largely associated
with the lower troposphere above the atmospheric boundary
layer.\cite{BrethertonPetersBack04} Validation of TMI rain rate
against space-borne precipitation radar (PR) at sub-daily time scales
in the tropical Western Pacific\cite{IkaiNakamura03} show TMI
overestimating rain rate but with an approximately linear relationship
to PR.  We have performed a number of checks to verify that results
are not substantially impacted by a high rain rate cutoff in the
algorithm ($25~$mm/h), including comparison to regions where cutoff
occurences are very low, such as the eastern Pacific (Fig. 1). The
clearest check is that the essential features are identical for the
cloud liquid water, whose measurement cutoff of $2.5~$mm is never
reached (see Supplementary Information).  
SST data here are averages over non-flagged neighbors in
space and time, since SST is not retrieved at high rain rates.
 
The critical value $w_c$ is determined by an iterative procedure with
an initial guess, followed by a fit to \eref{order} above $w_c$. Error
bars in Fig. 3 are standard errors of the variance $\sigma^2_P(w;L)$,
determined via the zeroth, second and fourth moments of $P(w)$.
Individual measurements of $P(w)$ are considered independent, which
holds well between satellite overpasses, though not within individual
tracks.
\bibliographystyle{prl}
\bibliography{./../../../../../bibliography/bibliography}

Supplementary Information is provided at
www.nature.com/nphys/journal/v2/n6/abs/nphys314.html

\begin{acknowledgments}
This work was supported under National Science Foundation grant
ATM-0082529 and National Oceanic and Atmospheric Administration grants
NA05OAR4310013 (JDN and OP) and the US Department of Energy
(W-7405-ENG-35) (OP).  We thank D. Sornette for connecting the
authors, and the RSS rain team for discussion.
\end{acknowledgments}
\end{document}